# Non-thermal influence of a weak microwave on nerve fiber activity

M. N. Shneider[1,*] and M. Pekker[2]

[1] Department of Mechanical and Aerospace Engineering, Princeton University, Princeton, NJ 08544, USA

[2] MM Solution, 6808 Walker Street, Philadelphia, PA 19135, USA

**Abstract**
This paper presents a short selective review of the non-thermal weak microwave field impact on a nerve fiber. The published results of recent experiments are reviewed and analyzed. The theory of the authors is presented, according to which there are strongly pronounced resonances in the range of about 30-300 GHz associated with the excitation of ultrasonic vibrations in the membrane as a result of interactions with the microwave radiation. These forced vibrations create acoustic pressure, which may lead to the redistribution of the protein transmembrane channels, thus changing the threshold of the action potential excitation in the axons of the neural network. The problem of surface charge on the bilayer lipid membrane of the nerve fiber is discussed. Various experiments for observing the effects considered are also discussed.

## I. Introduction

Studies of the effects of a weak (non-thermal) microwave radiation on the cellular structures began in the sixties-seventies of the last century with the pioneering work of Devyatkov et al in Soviet Union [1-3] and Fröhlich on the west [4-6]. In these pioneering studies have clearly shown that the reaction of cells to microwave radiation is highly selective. At certain frequencies the reaction of cells becomes extremely strong. Although various investigators reported different values of the resonant frequencies obtained in experiments, all these frequencies were in the range of tens of GHz [7,8] (see Figure 1 [8]).

The reason for the resonant interaction of the microwave with the cells was explained by either natural oscillations which are inherent to the cells themselves, as well as to separate molecular structures or individual molecules [1-6]. However, a clear understanding of the mechanism of interaction of microwave with the cells has not been achieved. As example, let us quote from a recent review [9]: "The energy of MiliMeterWave radiation is too low to directly disrupt any biochemical interaction, such as van der Waals or hydrogen bonds. Only a resonance-type interaction might lead to an appreciable biological effect. However, the existence of such a resonance at the cellular level is still unknown. According to the literature, cell membranes represent a promising potential target and this topic requires further experimental investigations in the coming years."

It should be noted that many experiments revealed the threshold character of the microwave influence. Starting with a certain minimum power the effects of radiation on the living organisms did not change, even if the power is increased by $10^5$ times and the local temperature increased by 0.1 degrees Celsius [8,10,11].

---

[*] m.n.shneider@gmail.com

As an example of stimulated cell activity, was presented in [8], in which the effect of microwave radiation on the synthesis of colicin E-coli C600(E1) was studied. It was shown that the effect of the microwave has a pronounced threshold, a resonance nature and depends on the exposure time. At a certain minimum intensity the saturation was reached: the impact of radiation on the organism did not change, even if the intensity was increased in ~1000 times.

Apparently, Fröhlich was the first to call attention to the coincidence of the resonance frequencies of microwave radiation with frequencies of natural oscillations of the cell membrane. Indeed, considering the cell membrane as thin plate of thickness $d_s = 9$ nm, in which the speed of sound $v_s \approx 1500$ m/s, then the frequency of $N^{th}$ acoustic harmonics in it is equal to:

$$f_N \approx \frac{v_s}{4d_s}(2N-1) = 41.6 \cdot 10^9 \cdot (2N-1), \text{ Hz.} \tag{1}$$

The exact value of the speed of sound in the membrane is not known, it is assumed to be close to the speed of sound in water [12,13]. Membrane thickness is also not equal for different nerve fibers and varies from 7 to 10nm [14, 15, 16]. In addition to purely transverse oscillations, a shear oscillations, corresponding to the azimuthal vibrations of thin-walled cylinder, can be excited in the membrane.

Fröhlich supposed in his works that the relationship between the natural oscillations of the membrane with microwave irradiation is carried out by proteins having a dipole point and "floating" in the membrane. He proposed a hypothesis of Bose condensation of sound waves in the membrane [4], which not been confirmed experimentally, since free oscillations at ultrasound frequencies are strongly damped. However, Fröhlich's theoretical papers attracted the attention of physicists to the problem of the interaction of microwaves with the membrane and stimulated the search for mechanisms of this interaction.

Another natural mechanism of interaction of microwave radiation with cellular membranes is the interaction of the electric field of the microwave with the charges sitting on the surfaces of the membrane. Therefore, the question of the charge of the membrane surface is key to understanding the interaction of cells with electromagnetic fields. Indeed, if the membrane is electrically neutral, the electromagnetic field can have only a negligible effect on the membrane due to the interaction with the dipole heads of phospholipids and proteins in it that have a dipole moment. If the membrane is charged and ions are bound to its surface strongly enough, the electrical component of the electromagnetic field will result in the membrane deformations. If the ions are bound to the surface is relatively weak, the ions on the membrane surface will move freely along its surface without causing any effect on it.

The question of the value of the surface charge of cell membranes has been studied in many experimental works. We will not go into all the papers, we refer only to [17] and [18], where the experimental data of the surface charge of cell membranes obtained by different methods; Range of obtained values of the surface charge density is quite broad: $\sigma_m = 0.3 - 0.002$ C/m$^2$. The review [18] states that an average of about 10% of lipids ape negatively charged.

In [13,19,20] Krasil'nikov considered several model problems related to the oscillations of a charged thin shell, that mimics the cell membrane located in a liquid electrolyte, under influence of the periodic electric field. The resonant frequency and Q-factor, corresponding to natural oscillations of the membrane of spherical vesicles are found in [21]. The obtained results were used for evaluation the displacement of the free ions in the vicinity of the shell. However, major questions were remained unanswered: what is the value of the surface charge of the membrane

(it was assumed to be given)? How strongly the surface charge is bounded with the membrane? And what the mechanism of action of weak electromagnetic fields on the cellular processes?

Recently, several experimental studies had been performed, which showed that a relatively weak microwave radiation with intensities of 0.01–1W/m2 in the range of 50–100 GHz leads to the spontaneous excitation of neural activity (see, for example, [17]). The interaction of microwave radiation with a nerve fiber was studied in [21] on the example of the mouse cerebral cortex, placed in the cerebrospinal fluid (CSF) solution and illuminated by plane-polarized microwaves at a frequency of 60.125 GHz. A periodic rectangular current pulse of a small amplitude was initiated in the nerve tissue sample to stimulate the action potential. With increasing of the microwave radiation intensity a significant increase in the frequency of the membrane self-excitation after the exposure to microwave radiation, as well as, the shortening of the recovery time of the resting membrane potential was observed in the experiments. The results obtained in [21] seem to be surprising. For example, at microwave intensity of 492 nW/cm$^2$, the frequency of the action potential self-excitation was higher than at the intensity of 71 nW / cm$^2$, but at the intensity of 737 nW/cm$^2$, the action potential did not excited at all. In other words, with the intensity of the microwave radiation increasing the threshold for the action potential excitation decreases, but at a certain critical value of the microwave intensity the action potential is blocked. This and other results cannot be explained in terms of thermal effect, since the intensities of the microwave are too small to change the temperature of the membrane and CSF solution. Note that, as shown below, the frequency of the microwave source selected in [21] is close to the resonant frequency of the natural oscillations of the membrane.

This raises the following question: what physical processes other than temperature may affect on the the threshold of the action potential excitation? In our opinion, such a reason could be the changing of the local surface density of transmembrane ion channels, in particular the sodium channels. It is known that these channels are not fixed in any specific locations of the membrane and can be displaced as a result of so-called lateral diffusion (the corresponding coefficients of lateral diffusion lie within $D_L = 10^{-13} - 10^{-14}$ m$^2$/s  [22, 23, 24]).

As shown in [25, 26] on the basis of the analysis of the equations of Hodgkin - Huxley, the surface density changing of sodium channels ($\propto g_{Na}/g_L$) decreases (increases) the threshold of the action potential (Figure 1). The values $g_{Na}$ and $g_L$ are, respectively, the maximum Na and leak membrane conductances.

At present time there are special actuality of studies exploring the microwave radiation impact on living organisms is associated with increasing levels of electromagnetic pollution of the human environment (related to the development of mobile communications, computers, radio, television, etc). The permissible standards of radiation power of electronic devices, depending on their frequency, been developed by joint efforts of scientists, engineers and physicians. An example of American standards of radiation safety at different frequencies [32] is shown in Figure 2. These standards are based on calculations of the thermal effects of microwave radiation on the human body. The allowable power of microwave radiation increases linearly in the range of 20 MHz - 20GHz and then remains constant, as can be seen from Figure 4. A plateau at frequencies above 20 GHz is due to the fact that at these frequencies the penetration ability of the microwave falls to the level of hundreds microns and radiation is almost completely absorbed in the skin [2].

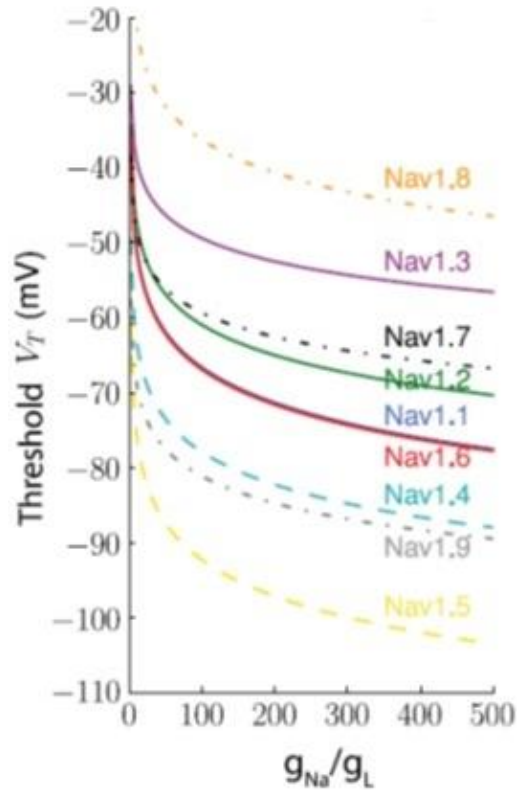

**Figure 1.** Influence of Na activation characteristics on spike threshold. Threshold as a function of the ratio $g_{Na}/g_L$ for the 9 types of voltage-gated sodium channels [27] with characteristics reported in (Angelino and Brenner, 2007 [28]). For each channel type, threshold obtained across the dataset is plotted. Nav1.[1,2,3,6] are expressed in the central nervous system, Nav1.[4,5] are expressed in cardiac and muscle cells and Nav1.[7,8,9] are expressed in the peripheral nervous system. Nav1.6 is expressed at the action potential initiation site [29–31] ([26] J. Platkiewicz and R. Brette, A Threshold Equation for Action Potential Initiation, PLoS Comput. Biol. **6**, e1000850, (2010)).

A few years ago, it has been proposed to increase the frequency of the electromagnetic radiation used for telecommunications up to 30–300 GHz. [33, 34]. This is based on the fact that at such frequencies, it is possible to transmit signals at a rate higher than 2 gigabits per second, which has enormous potential both for civilian purposes, as well as for the military. On the other hand, the increase in frequency makes it possible to use low-power energy sources that essentially reduce the electromagnetic radiation flux down to lower than 1 W/m$^2$. At such power of the incident radiation, the thermal effect on the nerve fiber is insignificant and cannot affect the vital functions of the human brain and nervous system. In 2013 the first Wi-Fi routers operating at a frequency of 60-80 GHz appeared on the market [35,36].

However, as noted above, there are resonance effects in addition to the thermal effect associated with the microwave absorption in the tissues of the body, (the nature of which, until recently, remained unclear). This can lead to physiological implications with much lower intensities than the permitted safety standards. These resonances are shown schematically in Figure 2 and highlighted with a question mark.

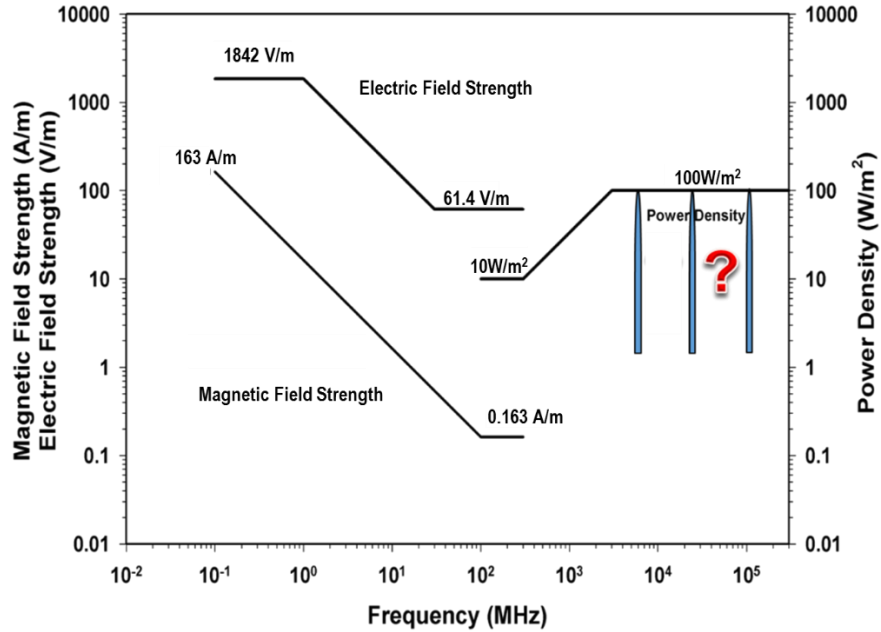

**Figure 2.** Acceptable safety standards of power electronic devices, depending on the frequency. Arrows indicate the resonant frequency observed in the experiments: 42.3, 53.8, and 61.2 GHz [39, 40] ([32] IEEE Standard for Safety Levels with Respect to Human Exposure to Radio Frequency Electromagnetic Fields, 3 kHz to 300 GHz (2006))

The dielectric constant of water depends on microwave frequency and is a complex function: $\varepsilon = \varepsilon' + i\varepsilon''$. Corresponding values of the refractive index $n$, the attenuation coefficient $\kappa$ and the attenuation length $a_d$ are determined by the expressions [37]

$$n = \sqrt{\left(\sqrt{\varepsilon'^2 + \varepsilon''^2} + \varepsilon'\right)/2}\,,\quad \kappa = \sqrt{\left(\sqrt{\varepsilon'^2 + \varepsilon''^2} - \varepsilon'\right)/2} \qquad (2)$$

$$a_d = \frac{c}{2\pi f \kappa}. \qquad (3)$$

The corresponding wavelength of the microwave in a medium $\lambda = \lambda_0/n$ ($\lambda_0 = c/f$, $c$ the wavelength and speed of light in vacuum). The amplitude of the electric field in the medium is attenuated as: $E(z) \sim \exp(-z/a_d)$, where z it the depth of penetration of a plane microwave incident normal to the surface. An example of dependencies on the frequency of the microwave of the real and imaginary parts of the dielectric constant, the refraction index and the attenuation coefficient $\kappa$ and the characteristic attenuation length $a_d$ in water at a temperature of $T = 25°$ C are shown in Figure 3. The calculations are made on the basis of the data of [38].

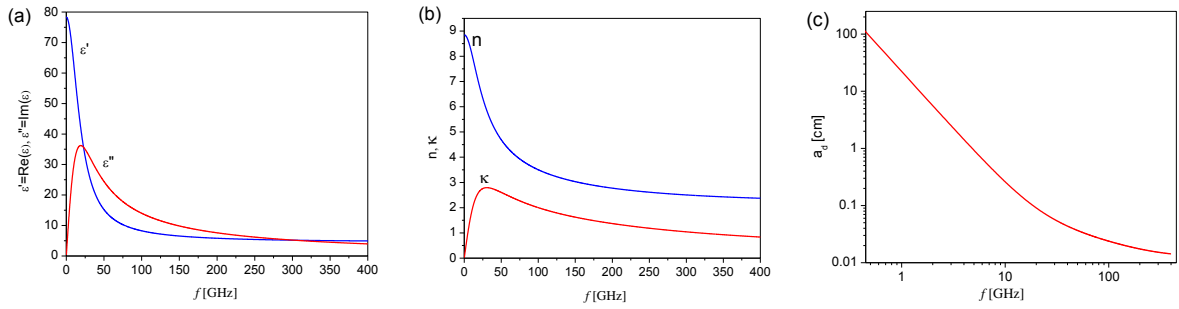

**Figure 3.** The dependencies on the microwave frequency: the real and imaginary dielectric constant values of water (a), the coefficient of refraction $n$ and attenuation $\kappa$ (b), and the attenuation length $a_d$ (c) at water temperature $T=25^0$ C.

Let us formulate the questions that we will try to respond to the main part of the article.

- How the microwave radiation leads to a redistribution of transmembrane ion channels? Why microwave interaction with membranes has a resonant character and what is the dependence of the characteristics of interaction on the intensity of the microwave radiation?
- What are the typical values of the surface charge of the membrane, and how strong the surface charges are bounded with the membrane?
- What are the biological consequences of redistribution of the surface density of the ion channels in the axon membrane?
- What experiments and theoretical studies should be done to make clearer the understanding of non-thermal interaction of low intensity microwaves with the nerve cells?

## II. Impact of low intensity microwave radiation on the neuron. Qualitative picture.

A schematic view of a neuron is presented in Figure 4. Arrow shows the direction of incidence of the microwave. Assume that the microwave wave is polarized such a way that electric field is directed along the axon. It is also shown myelin portion of the axon. The thickness of the membrane, as mentioned above, lies within 7-10 nm [14, 15, 16]. The length and diameter of the axons in different organisms and different elements of the nervous system varies widely. Thus the length of inhibitory interneuron axons is shorter than 1 mm, while the length of the peripheral nervous system axons may reach one meter and more [41]. The largest axons in the mammalian peripheral nervous system (PNS) reach a diameter of ~ 20 μm [42]. However, the greatest diameter has the squid giant axon with a diameter close to 1 mm [43], whereas the diameter of unmyelinated cortical axons in the mammalian brain varies between 0.08 and 0.4 μm [44, 45].

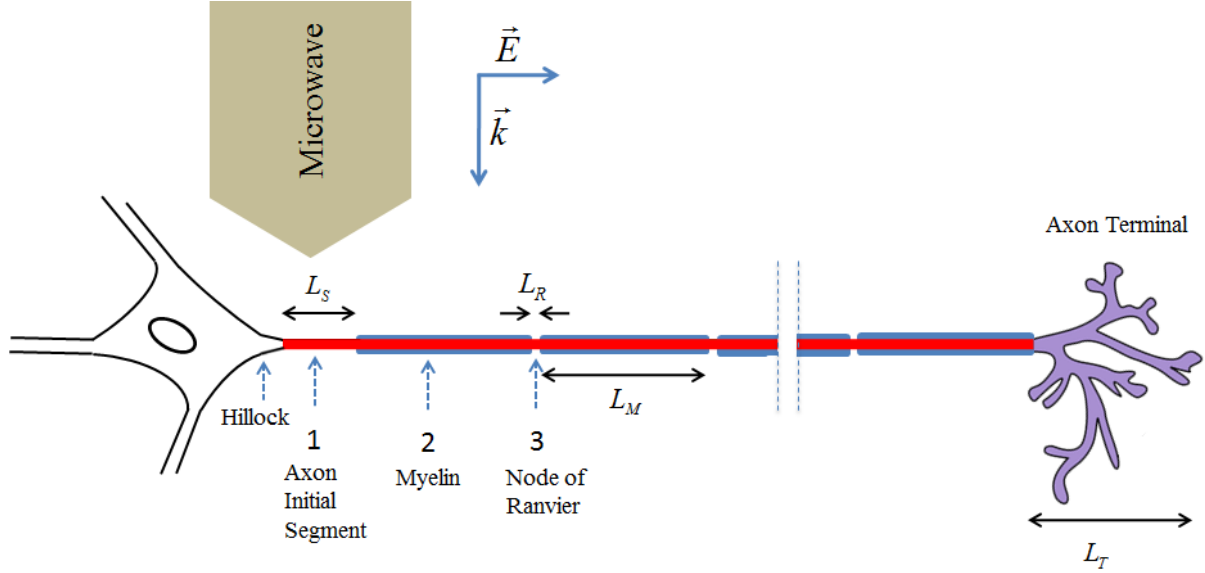

**Figure 4.** Scheme of a neuron with the characteristic elements of the myelinated axon.

$L_s \approx 10-50\,\mu m$, $L_R \approx 1-2.5\,\mu m$, $L_M \approx 100-1000\,\mu m$, $L_T \approx 10-2000\,\mu m$.

Initiation of the action potential (AP) is the subject of a large number of studies (see example t [46, 47] and references therein). At present it is generally accepted that the initiation in myelined fiber begins with the appearance of spikes in axon initial segment (AIS) (Figure 6), which lead to the excitation of the action potential in the axon. Typically the size of AIS is in the range 20-50 μm, but it can reach 75 μm and even, 200μm. To initiate an action potential, it is necessary that the amplitude of the spike was higher than the excitation threshold, ie ~ 5-10 mV. The amplitude of the voltage required for the initiation of the AP lower for a spike with longer time duration. In general, the conditions for the excitation of the action potential depend on the length of the AIS and the distribution of the surface density of transmembrane ion channels [46, 47].

Typical dimensions of AIS for all types of axons are much smaller than the microwave wavelength in the range of tens-hundreds GHz, so we can assume that the electric field acting on the charged surface of the AIS membrane, uniform in space and varying only in time. (The surface charge of the membrane is discussed in Section IV, [48]). Since the AIS is bordered at one end by hillock and at the other by a myelin sheath (Figure 4), we can consider the charged AIS membrane as a plate clamped on the ends (Figure 5). This is valid because the membrane thickness is several orders of magnitude thinner than the diameter of the thinnest neuron. Displacements of the membrane in the case when the charges on the inner and outer sides of the membrane have different signs are shown in Figure 5a; and the case when charges of surfaces have the same sign is shown in Figure 5b. Forced oscillations of a charged elastic plate in an alternating electric field have pronounced resonances. Corresponding characteristic resonant frequencies are $\omega_r \approx \pi p \dfrac{c_l}{h}$, where $p = 1,2,...$ (Part V, [49])

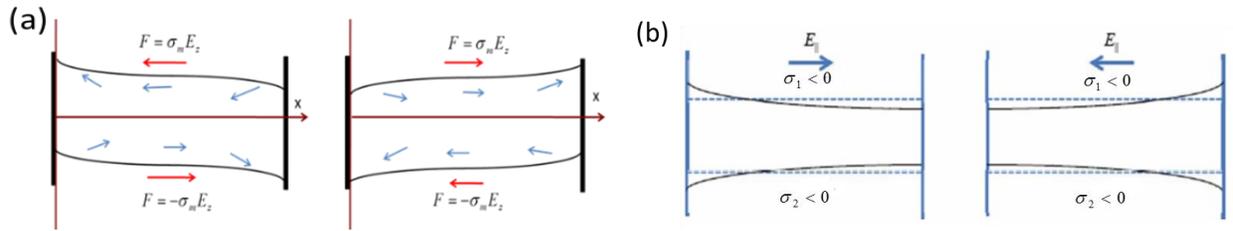

**Figure 5.** Membrane deformations causes by the microwave electric field directed. The edges of the membrane (unlimited in *y* direction) are clamped. The blue arrows indicate the elastic displacements inside the membrane. (a) corresponds to the charges of opposite sign on the membrane surfaces, (b) – the same sign.

Forced longitudinal vibrations of the plate-membrane lead to a redistribution of transmembrane proteins (ion channels) due to acoustic radiation pressure (see Section III [49, 50]) In the case of excitation of the longitudinal standing acoustic wave in the membrane, ion channels are shifted to the edges and a rarefied area is formed at the center (Figure 6).

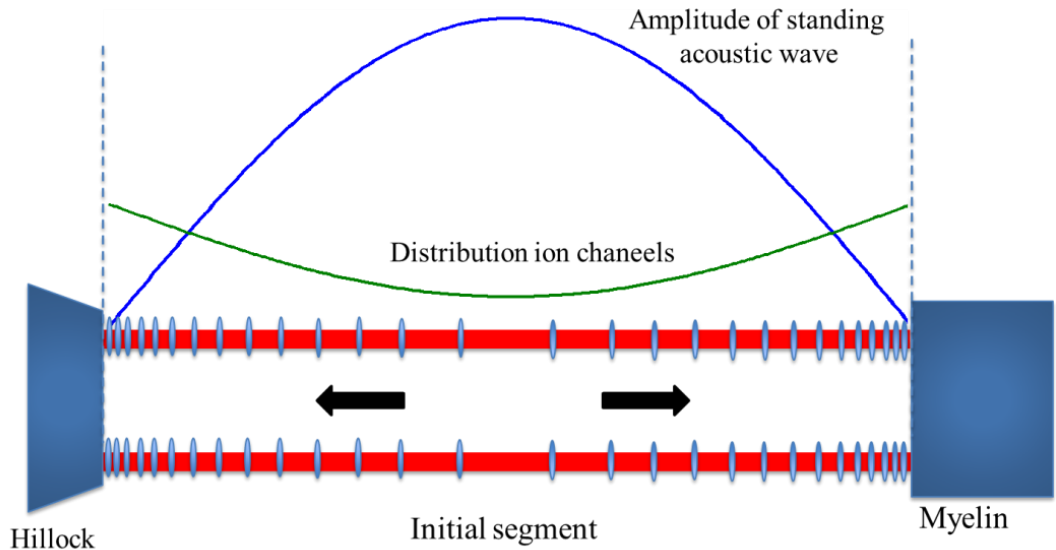

**Figure 6.** An example of the transmembrane ion channels redistribution under the influence of the acoustic radiation pressure. The arrows show the direction of the forces acting on ion channels protein. It is assumed that the protein density is close to the density of the membrane

The redistribution of sodium transmembrane channels caused by longitudinal acoustic vibrations leads to a decrease in the excitation threshold of the action potential (or spike) in the areas with enhanced surface density of ion channels. And, with a significant rarefaction and sufficiently long rarefied region is possible to block the action potential propagation (Section IV [50]).

Note that since the dimensions of non-myelined axon terminals (AT) can reach several tens of microns, there also may be a significant redistribution of protein channels, which also can lead to physiological effects.

## III. Lateral diffusion and drift of transmembrane proteins in an acoustic field [49, 50]

In this section we will try to answer questions about the reason of redistribution transmembrane channels and how long it takes.

We will not consider numerous experimental works related to the diffusion of transmembrane proteins in biological membranes and will restrict ourselves to a statement of fact that the transmembrane proteins not fixed and are subject to the so-called lateral diffusion. Typical values of the lateral diffusion coefficient are within $D_L = 10^{-13} - 10^{-14} m^2/s$ [22, 23, 24].

A well-known fact in acoustics that suspended particles in the field of a traveling wave are displaced either to the region of maximum intensity of the acoustic oscillations or to the minimum intensity. In the field of the standing wave suspended particles are collected in the antinodes in the nodes regions [51, 52]. The direction of the suspended particles displacement is determined by their relative density with the density of the medium.

We will consider transmembrane proteins in the membrane as suspended particles in the mediumin which there can be excited the acoustic vibrations. For simplicity of analysis and without loss of generality, we will assume the proteins as spherical particles of radius $a = d/2$ ($d$ is the thickness of the membrane) and having a density approximately equal to the density of the membrane $\rho \approx 900 \text{ kg/m}^3$. According to [53] the forces $F_t$ and $F_s$ acting on the suspended spherical particles by a plane traveling sound wave

$$F_t = \frac{4\pi}{9} a^6 k^4 \overline{E} \qquad (4)$$

or, in a plane standing sound wave:

$$F_s = \frac{2\pi}{3} \cdot a^3 k_n \overline{E} \sin(2k_n x) \qquad (5)$$

Here, $\overline{E} = \frac{1}{2}\rho\omega^2 A^2$ is the energy density in the wave; $\omega = 2\pi f$ is the angular frequency of ultrasound, $A$ is the amplitude of the oscillatory displacements in the wave; $k = 2\pi/\lambda = c_s/\omega$ is the wave number; B (5) $x$ is the position of a particle in the standing wave, $k_n = \frac{n\pi}{l}$, $n=1, 2, $, $l$ is the size of the area where a standing wave exists.

An initial homogeneous distribution of the transmembrane proteins is disturbing in the field of the acoustic wave. The transmembrane proteins will drift under the influence of the acoustic radiation pressure until their forced drift will not compensates by the reverse diffusion flux. The continuity equation for the density $n_{ch}$ of protein per unit surface area of the membrane in the acoustic wave field is:

$$\frac{\partial n_{ch}}{\partial t} + \nabla \cdot \mathbf{\Gamma} = 0; \quad \mathbf{\Gamma} = \mu_L n_{ch} \mathbf{F} - D_L \nabla n_p \qquad (6)$$

Here $\mathbf{F}$ either $F_s$ or $F_t$, $\mu_L$ is the lateral mobility of transmembrane protein channels, which is associated with the lateral diffusion and the local temperature by the Einstein relation $\mu_L = D_L/k_B T$.

As was shown in [49] (see Section IV), that the most favorable areas for the excitation of longitudinal vibrations in the membrane in the myelined nerve fiber are AIS and AT (Figure 4), since the amplitude of these oscillations $\sim 1/l^2$, where $l$ is the length of non-myelined area. Therefore, we can assume that all the results obtained below are related to the AIS or AT.

Assuming that the external force **F** is oriented along the axis of the axon and taking into account that the membrane thickness is much smaller than the radius of the axon, the equation (6) for the density of the transmembrane channels can be rewritten as:

$$\frac{\partial n_{ch}}{\partial t} + \frac{\partial}{\partial x}\left(-D_L \frac{\partial}{\partial x} n_{ch} + \mu_L F n_{ch}\right) = 0. \tag{7}$$

The steady-state distribution of the protein channels density satisfies the Boltzmann distribution:

$$n_{ch} = n_0 \exp(-U/k_B T) \tag{8}$$

with the potential

$$U(x) = -\int_0^x F(x) dx . \tag{9}$$

The factor $n_0$ can be easily found from the condition of conservation of the total number of ion channels along the length of the membrane.

$$n_c(x) = \frac{n_{c,t0}}{\frac{1}{l}\int_0^l e^{-U(x)/k_B T} dx} e^{-U(x)/k_B T} , \tag{10}$$

where $n_{ch,0}$ is the undisturbed density of ion channels per unit membrane surface.

The dimensions of the initial segment $L_s$ vary between 10-50 μm, which is much smaller than the wavelength of ultrasonic vibrations. Since the initial segment on the one side is bordered with the myelin covered area, and on the other - with hillock, the longitudinal displacement of the ends of the initial segment membrane is suppressed. Thus, only standing waves with wave numbers $k_n = \pi n / L_s$, $n = 1,2,3\ldots$ can exist in the membrane of the initial segment (Figure 6).

For the case of a standing wave from the formula (10) follows:

$$\frac{n_{s,ch}}{n_{ch,0}} = \frac{e^{-\frac{\pi}{3}a^3 \overline{E}_s (1-\cos 2k_n x)/k_B T}}{\frac{1}{L_s}\int_0^{L_s} e^{-\frac{\pi}{3}\cdot a^3 \overline{E}_s (1-\cos 2k_n x)/k_B T} dx} = \frac{e^{\xi_s \cos(2k_n x)}}{I_0(\xi_s)}, \quad \xi_s = \frac{\pi}{3}\cdot\frac{a^3 \overline{E}_s}{k_B T} , \tag{11}$$

where $n_{s,ch}$ is the density of the ion channels in the case of the standing ultrasound wave, $\overline{E}_s = \frac{1}{2}\rho_0\omega^2 A_s^2$ is the energy density of the standing ultrasonic wave, and $I_0$ is the modified Bessel function [54]. Figure 7 shows the dependence of the equilibrium density of sodium channels along the initial segment.

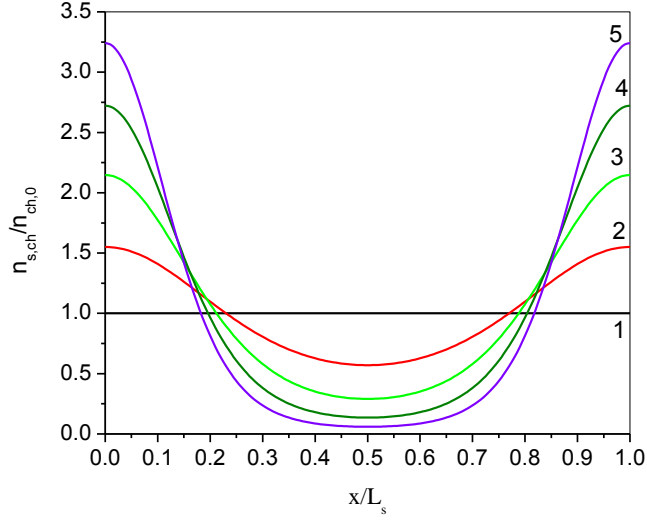

**Figure 7.** The relative densities of sodium channels along the initial segment interacting with the main mode of ultrasound standing wave ($n=1$) for different values of the parameter. 1 – $\xi_s = 0$, 2 – $\xi_s = 0.5$, 3 – $\xi_s = 1$, 4 – $\xi_s = 1.5$, 5 – $\xi_s = 2$.

The difference between the maximum and minimum densities of ion channels referred to the average density of the channels in the membrane is:

$$\frac{\Delta n_{s,ch}}{n_{ch,0}} = \frac{2\sinh(\xi_s)}{I_0(\xi_s)} \qquad (12)$$

Note that, at $\xi_s > 1$, the sodium channels are almost completely shifted to the edges of the initial segment. On the one hand, this may lead to the blockage of the action potential. On the other hand, this may reduce the excitation threshold of the action potential, as there are areas in which the channel density is much higher than at the unperturbed equilibrium. Examples of forced redistribution of transmembrane channels which lead to the blockage of the action potential and to self-excitation are considered in the following sections.

Figure 8 shows the density of sodium channels at different moments in time obtained through numerical solution of Eq. (7). It can be seen that the channel redistribution happens quite quickly. At $\xi_s = 2$, it takes approximately 5% of the lateral diffusion characteristic time, $t_0 \approx L_s^2 / D_L$.

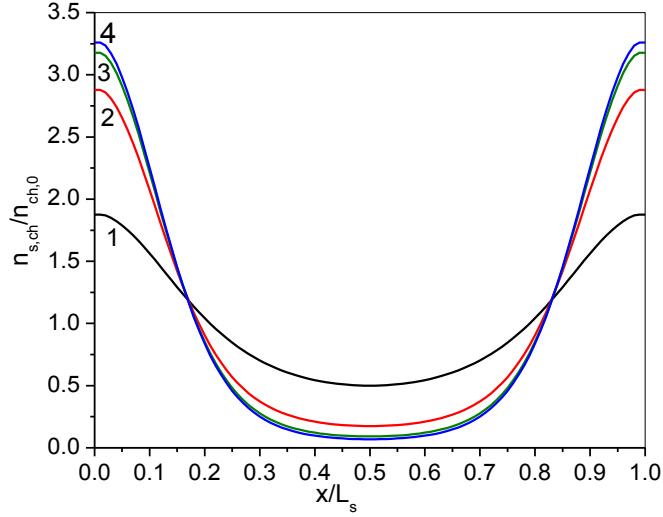

**Figure 8.** The relative densities of sodium channels along the initial segment interacting with the main mode of ultrasound standing wave ($n=1$) for parameter $\xi_s = 2$ at different moments in time. 1 corresponds to $t/t_0 = 0.01$, ($t_0 = L_s^2/D_L$), 2 − $t/t_0 = 0.03$, 3 − $t/t_0 = 0.05$, 4 − $t/t_0 = 0.07$.

It follows from Figure 7 that for $\xi_s = 2$, the formation of a new equilibrium distribution of ion channels happens very quickly in the time interval of order $0.03 t_0$. In other words, the initial formation stage of the regions with reduced or increased density of ion channels is much shorter than the channel density restoration time after turning off the ultrasound.

For the case of a traveling wave, we must take into account that the density of energy of the acoustic waves is decaying due to the viscosity as

$$\bar{E}_t = \bar{E}_{t,0} \exp(-2x/a_w), \tag{13}$$

where $a_w = c_s^3/(\omega^2 \nu)$ is the attenuation length of sound waves, $\nu \approx \nu_0/(1+\omega^2 \tau_\nu)$ is the kinematic viscosity in water [55], $\nu_0 = 10^{-6}\,\text{m}^2/\text{s}$, $\tau_\nu \approx 8 \cdot 10^{-12}\,\text{s}$ [56], $c_s \approx 1450\,\text{m/s}$ is assumed speed of sound in the membrane. Figures 9a and 9b show the dependencies of the attenuation length $a_w$ and kinematic viscosity $\nu$ on the sound frequency in water.

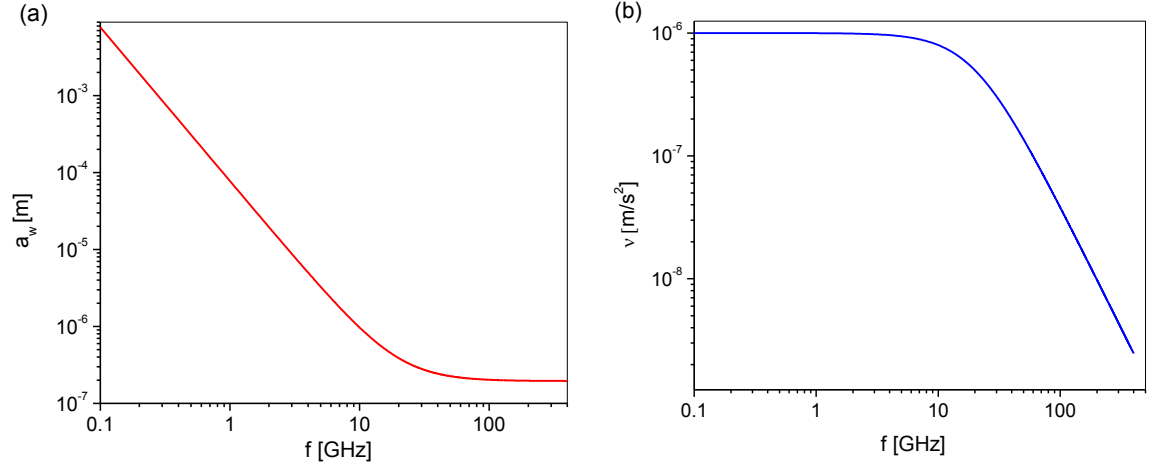

**Figure 9.** Sound attenuation length (a) and viscosity (b) versus frequency in water

## IV. Forced vibrations of the membrane in a microwave field [49]

Geometrical and mechanical properties of nodes Ranvier, the initial segment, are very different from the terminal region and the areas covered by the myelin (Fig. 4). Therefore, their vibrations (longitudinal and transversal), arising due to the action of the electric component of the electromagnetic field can be treated independently, assuming that the ends of the vibrating areas are fixed. Since the thickness of the axon membrane in the nerve fiber is much smaller than its radius, we will consider each part of axon as a plate, unlimited in the $y$-direction, with uniformly distributed surface charge $\sigma_m$ [C/m$^2$], which is firmly fixed to the membrane surface. A key issue how rigidly ions are fixed on the membrane surface and what is the value of the surface charge will be briefly discussed in Section V.

As mentioned earlier, the wavelength of the electromagnetic wave in a medium is equal to $\lambda = \lambda_0 / n$. In the frequency range $f \sim 50\text{-}100$ GHz, коэфициент преломления $n$ лежит в интервале 5-3 (the frequency dependence $n(f)$ is shown in Figure 5) and, respectively, $\lambda(f)$ is in the range 1-2 mm. That is, the micriowave wavelength is much greater than the length of any element of the myelined nerve fiber (shown in Figure 6) Therefore, we can assume the electric field acting on the charged membrane is spatially uniform and varies only in time. Figure 7a shows the displacement of the membrane in the case where the charges on the inner and outer sides of the membrane have a different sign, 7b – a sign of the surface charges is the same. The force acting per unit area of the charged surface of the membrane is equal:

$$\mathbf{F} = \sigma_m \mathbf{E} \tag{14}$$

Here $\mathbf{E}$ is the electric field of the microwave acting on the surface charges, $\sigma_m$ is the surface charge density. For the electromagnetic wave of the intensity $I_{MW}$, the amplitude of the electric field in the medium [57]:

$$|\mathbf{E}| = \sqrt{\frac{2I_{MW}}{\varepsilon_0 n c}} = \frac{E_{0,a}}{n^{1/2}}. \tag{15}$$

Here, $I_{MW}$ is the microwave intensity in the vicinity of the membrane and $E_0$ is the corresponding amplitude of the electric field in vacuum. In the frequency range of 50-100 GHz, $n^{1/2} \approx 2$

(Figure 3b), i.e., the electric field acting on the membrane, is only half than the electric field in vacuum. Since, as shown in [47], the normal to the membrane component of the electric field practically does not contribute to the displacement of transmembrane proteins, we will consider only the case when the microwave electric field component is directed along the axon. Hereinafter, for simplicity, we will assume that $n^2 \approx \varepsilon \approx \varepsilon'$. It does not lead to a significant error in the frequency range of our interest (Figure3b).

The system of equations describing the longitudinal vibrations of a membrane, such an elastic bar under the influence of external driving forces, have the form [58]:

$$\frac{\partial^2 u_x}{\partial t^2} = c_l^2 \frac{\partial^2 u_x}{\partial x^2} + c_t^2 \frac{\partial^2 u_x}{\partial z^2} + \nu \frac{\partial}{\partial t}\left(\frac{\partial^2 u_x}{\partial x^2}\right) + \left(c_l^2 - c_t^2\right)\frac{\partial^2 u_z}{\partial z \partial x}$$
$$\frac{\partial^2 u_z}{\partial t^2} = c_t^2 \frac{\partial^2 u_z}{\partial x^2} + c_l^2 \frac{\partial^2 u_z}{\partial z^2} + \nu \frac{\partial}{\partial t}\left(\frac{\partial^2 u_z}{\partial x^2}\right) + \left(c_l^2 - c_t^2\right)\frac{\partial^2 u_x}{\partial z \partial x}$$
(16)

Here $u_x, u_z$ are displacements along axes x and z; $c_l$, $c_t$ are the longitudinal (along the *x*-axis) and transverse (along the *z*-axis) speed of sound. In our model, we assume that $c_l \approx c_t$.

Since, as we assumed, that the charges are only on the surface of the membrane, there are no terms associated with the electric field in the equations (16). The boundary conditions corresponding to the longitudinal electric field, in the case where the surface charges on the inner and outer surfaces of the membrane $\sigma_m$ are equal in absolute value but are opposite in sign (Figure 5a):

$$\left.\frac{\partial u_x}{\partial z}\right|_{z=h} = \frac{F_x}{c_l^2 \rho_0}$$
$$u_x|_{z=0} = 0 \quad ,$$
$$u_x|_{x=0, x=L} = 0$$
(17)

and, when the sign of the surface charge of are equal (Figure 5b):

$$\left.\frac{\partial u_x}{\partial z}\right|_{z=h} = \frac{F_x}{c_l^2 \rho_0}$$
$$\left.\frac{\partial u_x}{\partial z}\right|_{z=0} = 0 \quad .$$
$$u_x|_{x=0, x=L} = 0$$
(18)

Here $F_x$ is the force tangential to the membrane surface

$$F_x = \sigma_m E_{II} = \sigma_m \sqrt{\frac{2I_{MW}}{\varepsilon_0 nc}} e^{i\omega t} = \sigma_m \frac{E_{0,a}}{\varepsilon^{1/4}} e^{i\omega t}$$
(19)

It is not obvious what value should be substituted in (17). If the surface charge is on the surface of the membrane, it should be taken $\varepsilon = n^2$, but if it is "deepened" into the membrane then $\varepsilon = \varepsilon_m = 2$ [59].

We will look for a solution of equations (14), in accordance with the boundary conditions (17), in the form:

$$u_x = A_{II} e^{i\omega t} \sin(k_n x)\sin(\xi z)$$
$$u_z = A_{\perp} e^{i\omega t} \cos(k_n x)\cos(\xi z) , \qquad (20)$$
$$k_n = \frac{\pi n}{L}$$

and, in the case of boundary conditions (18), in the form:

$$u_x = A_{II} e^{i\omega t} \sin(k_n x)\cos(\xi z)$$
$$u_z = A_{\perp} e^{i\omega t} \cos(k_n x)\sin(\xi z) , \qquad (21)$$
$$k_n = \frac{\pi n}{L}$$

Substituting (20) and (21) into (16) we obtain the expressions:

$$\left(\omega^2 - k_n^2 c_t^2 - i\omega v k_n^2 - c_t^2 \xi^2\right) A_{II} + k_n \left(c_l^2 - c_t^2\right) \xi A_{\perp} = 0 \qquad (22)$$
$$\left(\omega^2 - k_n^2 c_t^2 - i\omega v k_n^2 - c_l^2 \xi^2\right) A_{\perp} + k_n \left(c_l^2 - c_t^2\right) \xi A_{II} = 0$$

Substituting (20) into (17) we obtain the amplitude for the shear vibrations (Figure 5a):

$$A_{II} = \frac{F_{n,x}}{c_t^2 \rho_0 \xi \cos(\xi h)} \qquad (23)$$

Substituting (23) into (22) and taking into account that we are interested in the frequencies $\omega \gg k_n c_l, k_n c_t$,

$$\xi^2 \approx \left(\omega^2 - i\omega v k_n^2\right) / c_l^2 \qquad (24)$$

we obtain:

$$A_{r,II} = \frac{1}{c_l \omega \cos(\xi h)} \frac{F_{n,x}}{\rho_0} , \qquad (25)$$
$$A_{r,\perp} = -\frac{c_l k_n}{\omega_r} A_{r,II}$$

Taking into account that

$$|\cos(\xi h)| = \left|\cos\left(\frac{\omega h}{c_l} + i\frac{v k^2 h}{2 c_l}\right)\right| = \sqrt{\cos^2\left(\frac{\omega h}{c_l}\right) + sh^2\left(\frac{v k^2 h}{2 c_l}\right)}, \qquad (26)$$

we can find from (26) the frequencies and the widths of resonances.

Since $\dfrac{vk^2h}{2c_l} \ll 1$, the resonant cyclic frequencies are

$$\omega_r = \left(\dfrac{\pi}{2} + \pi p\right)\dfrac{c_l}{h}, \quad p=0,1,2,..., \qquad (27)$$

and the width of the resonances is:

$$\delta\omega \approx vk_n^2/2. \qquad (28)$$

Substituting $\omega_r$ in (25), we find the resonant amplitudes $A_{II}$ and $A_\perp$

$$A_{r,II} = \dfrac{2}{\omega_r vk_n^2 h}\dfrac{F_{n,x}}{\rho_m}$$

$$A_{r,\perp} = -\dfrac{c_l k_n}{\omega_r} A_{r,II} \qquad (29)$$

Similarly, the resonant frequencies for the boundary conditions (18) (vibrations in Figure 5b):

$$\omega_{*r} = (\pi + \pi p)\dfrac{c_l}{h}, p=0,1,2,... \qquad (30)$$

and the vibration amplitudes:

$$A_{*r,II} = \dfrac{2}{\omega_{*r} vk_n^2 h}\dfrac{F_{n,x}}{\rho_0} = \dfrac{2}{\omega_{*r} vk_n^2 h}\dfrac{F_{n,x}}{\rho_0}$$

$$A_{*r,\perp} = -\dfrac{c_l k_n}{\omega_{*r}} A_{r,II} \qquad (31)$$

The resonance width for the vibrations shown in Figure 7b is the same as for the case shown in Figure 5a.

Let us choose for example the following parameters of the lipid membrane of the nerve fibers: $d = 2h = 10$ nm; $c_l \approx c_t \approx 1450$ m/s. Then, the resonant frequencies corresponding to vibrations shown in Figure 5a, are

$$f_{r,p} = \dfrac{\omega_r}{2\pi} = 75(1+2p)\,GHz, \qquad (32)$$

and, in Figure 5b:

$$f_{*r,p} = \dfrac{\omega_{*r}}{2\pi} = 150(1+p)\,GHz. \qquad (33)$$

It should be noted that due to the existence of natural statistical variations of geometrical parameters of the nerve fiber regions - their length and the membrane thickness - the response of the ensemble of neurons on the microwave will have a much broader spectral range of in the vicinity resonances. Let us estimate a spectral width of the resonance for the initial segment. From (22) it follows:

$$\omega^2 = c_t^2 \xi^2 + k^2 c_l^2. \tag{34}$$

Since for the vibrations shown in Figure 5a, the minimum value $\xi = \dfrac{\pi}{2h}$, and for the vibrations (Figure 5b) - $\xi = \dfrac{\pi}{h}$, then, at $c_t \approx c_l$ it follows from the dispersion equation (34)

$$f_{r,0} = \frac{c_l}{4h}\left(1 + 2\frac{h^2}{L_S^2}\right) \tag{35}$$

$$f_{*r,0} = \frac{c_l}{2h}\left(1 + \frac{h^2}{2L_S^2}\right) \tag{36}$$

If there are many neurons in the sample with the lengths of the axon initial segments varying between 10-50 μm, then $\Delta f_r \approx 22\,\text{kHz}$ и $\Delta f_{*r} \approx 5.5\,\text{kHz}$. In fact, the resonance can be even more broadened if we consider subharmonics $k_n = n\pi/l$, and also that the thickness of the membrane may vary within the limits $6-10\,\text{nm}$.

Assuming $\bar{E} = \dfrac{1}{2}\rho\omega_r^2 A_{\parallel}^2$, and $a = h = d/2$, from (27), (29) and (5), the force acting on ion channels by the forced longitudinal vibrations of the membrane:

$$F_s \cong \frac{4\pi}{3}\frac{h}{k_n^3 v^2}\sigma_m^2 \frac{E_{0,a}^2}{\rho\varepsilon_m^{1/2}}\sin(2k_n x) \tag{37}$$

For the first resonance $k_1 = \pi/L_s$, the relative difference between maximum and minimum densities of the transmembrane ion channels $\xi_s$, defined by (11):

$$\xi_s \approx \alpha_s h\sigma_m^2 L_s^4 E_{0.a}^2 = 2\alpha_s h\sigma_m^2 L_s^4 \frac{I_{MW}}{\varepsilon_0 \varepsilon_m^{1/2} c}$$

$$\alpha_s = \frac{2}{3\pi^3 \rho_0 v^2 k_B T} \tag{38}$$

The microwave radiation, apparently, does not affect the nodes of Ranvier because $\xi_s$ depends on the length of non-myelined area in the fourth degree.

Let us estimate the redistribution time under the influence of the longitudinal force $F_s$, taking into account the lateral mobility $\mu_L = D_L/k_B T$:

$$\Delta t_s \approx \frac{L_s}{2V_d} \approx \frac{L_s}{2D_L}\frac{k_B T}{F_s} \approx \frac{1}{8\alpha_s D_L h}\frac{\varepsilon_0 \varepsilon_m^{1/2} c}{\sigma_m^2 L_s^2 I_{MW}} \quad (39)$$

It is assumed that the surface charges through which the microwave interacts with the membrane a firmly bonded to the surface. The value of the surface charge density is a critical parameter of the problem. In the simplest model case, considering the charged membrane as a planar capacitor with the potential difference equal to the rest potential $\Delta U_m \approx -70\,\text{mV}$, for the membrane thickness $d = 10\,\text{nm}$, the surface charge density can be estimated as $\sigma_m = \varepsilon_0 \varepsilon_m |\Delta U_m|/d \approx 1.24 \cdot 10^{-4}\,\text{C/m}^2$. The charges on surfaces of the membrane are considered equal in magnitude and opposite in sign. In this case, under the action of microwave radiation, the mechanical vibrations occurs in the membrane, such as shown in Figure 5a.

Table 1 shows the calculated values $\Delta n_{s,ch}/\langle n_{ch,0}\rangle$ and $\Delta t_s$ at different microwave intensities for the initial segment $L_s = 50\,\mu\text{m}$, $\sigma_m = 1.24 \cdot 10^{-4}\,\text{C/m}^2$ and the kinematic viscosity $\nu$ (Figure 9b) corresponding to the resonant frequency.

**Table1.** The calculated values $\Delta n_{s,ch}/\langle n_{ch,0}\rangle$ and $\Delta t_s$ at different $I_{MW}$.

| $I_{MW}\,[\text{W/m}^2]$ | $\Delta n_{s,ch}/\langle n_{ch,0}\rangle$ | $\Delta t_s$ [minutes] |
|---|---|---|
| 0.1 | 7.44 10$^{-2}$ | 4.45 10$^2$ |
| 1 | 7.35 10$^{-1}$ | 4.45 10$^1$ |
| 10 | 4.64 | 4.45 |
| 100 | 15.2 | 0.445 |

Note that the results shown in Table 1 represent the lower estimate of the effect. For a more realistic surface charge density (see next section), it can be expected that a significant redistribution of ion channels occurs more rapidly and at a much lower intensity of the microwave.

## V. The surface charge of the phospholipid membrane [48]

The problem of spatial distribution of electrical field in the vicinity of the biological membranes surface has been considered in many теоретических papers (see, example [60-65]). In all these works, the near-surface potential of the membrane was considered under the Gouy-Chapman theory [66, 67] or its later modification by Stern [68], in which the charge on the membrane surfaces is considered to be given. In these theories, the membrane was considered as a continuous dielectric, without taking into account its fine structure, and a surface charge was determined on the basis of the electrochemical properties of the dielectric surface (see, for example, [69, 70]).

In fact, It is known that the phospholipid molecules of cell membrane are forming a mosaic (matrix) structure in which dipole heads are directed towards the liquid (positively charged head faces outward membrane) [71]. The average surface area per molecule of the lipid is $\approx 0.5\,\text{nm}^2$, the length of the polar head is $\sim 0.5 - 1$ nm, the radius of the head is $\sim 0.2 - 0.3$ nm, and the distance between the hydrophilic heads of the membrane is in the range of 5-7 nm [15, 16]. The dipole moment of the phospholipid head is 18.5-25 D [72] (1 D = $3.34 \cdot 10^{-30}$ C·m), i.e.

more than 10 times greater than the dipole moment of water molecules. On the basis of geometrical dimensions of the cell membrane and the size of water molecules, it can be concluded that the free space between the head does not exceed the size of a water molecule (~ 0.2nm). That is, the membrane, interacting with the ions of surrounding liquid, cannot be considered as a dielectric medium with an infinitesimal dipoles size.

These facts allow the consideration of the following simplified model of the ion interaction with the membrane:
1. The membrane represents a matrix (Figure 10) with a mesh size $a \times a$. In the nodes of cells the dipoles are located; the dipole charge is q; the distance between the charges (the dipole length), d; the distance between the dipoles along the axis $z$, $l$.
2. An ion is a classical particle and cannot approach a dipole at a distance less than the sum of the radii of the head and the size of the ion. It is important that the ion can approach the membrane dipole heads close enough that would be "captured" by the potential well. This is a standard assumption in the theory of the interaction of ions with the surface of the dielectric [68]. Since the dipole moment of water is 10 times less than the dipole moment of the phospholipid head molecule and near the head cannot be more than one-two water molecules, the interaction between the ions located near the surface of the membrane and the water molecules can be neglected, as compared with the interaction of the ions with the dipoles of the membrane.

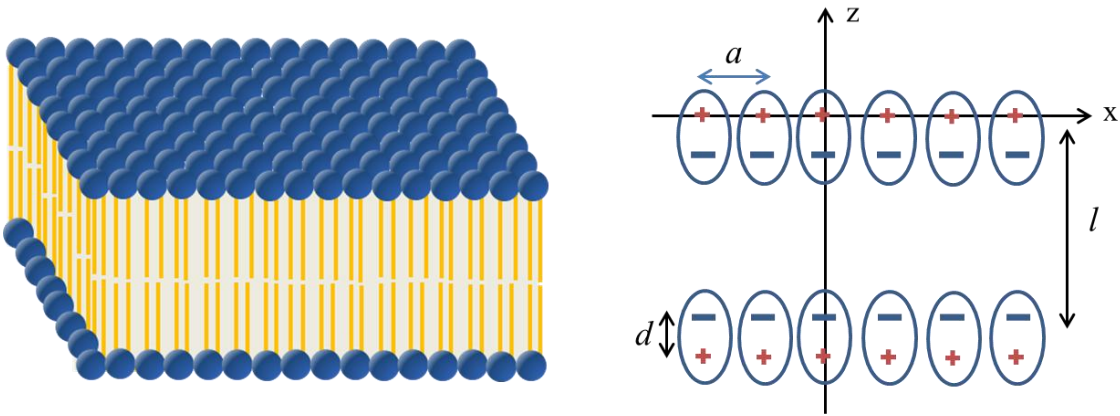

**Figure 10**. A simplified mosaic model of a membrane: the membrane is a matrix with dipoles in the nodes. *a* is the mesh size of the matrix, *d* is the distance between the charges of the dipole head, *l* is the distance between the dipoles along the z axis. .

The model of the phospholipid membrane (Figure 10) in saline was considered in [46], on the basis on the assumptions 1, 2,. Issues of the passage of ions through the cell membrane was not considered in [46], beqause the main question the authors were interested in: how strongly the surface charges are connected to the membrane, and what is the corresponding bounded surface charge density.It was shown that ions are firmly bounded to the surface of the membrane, the membrane surface is negatively charged, and the binding energy of the ions with the membrane $U$ is essentially higher than the thermal, $U/k_B T >> 1$. Therefore, the electric component of the microwave field, interacting with ions, transfers energy and momentum directly to the membrane. This interaction leads to forced mechanical vibration of the membrane and, as a result, to a redistribution of transmembrane protein ionic channels.

The compositions of negative ions inside and outside of the axon are different. The main negative ions inside are anion groups of macromolecules and phosphates, and the outside - the chlorine ions. Therefore, the binding energy of the ion with the membrane and the surface charge density will be different for the inner and outer surfaces of the membrane.

A self-consistent theory of the potential near the phospholipid membrane was developed in [48]. On the basis of this theory the surface charge density on the membrane was calculated. The results [48] e in agreement with the experimental data presented in [17, 18], according to which the surface charge density on the membrane $|\sigma_m| \sim 0.3 - 0.002$ C/m$^2$. Thus, the surface charge density on both sides of the membrane significantly exceed their difference that determines the resting potential, i.e. $|\Delta\sigma_m| \sim 10^{-4}$ C/m$^2$ $<< |\sigma_1|, |\sigma_2|$. Figure 11 shows the potential distribution inside and outside the membrane where the inner and outer sides of the membrane have different surface charges. Apparently, the forced vibrations of the membrane under the influence of the electric component of the microwave are corresponding to Figure 5b and the possible effect of the microwave is much stronger than presented in Table 1.

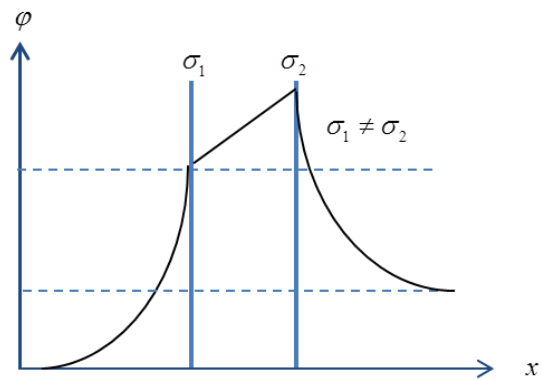

**Figure 11.** Potential distribution inside and outside the membrane when the inner and outer sides of the membrane have different surface charges.

In [73,74], the distribution of ions near the surface of the phospholipid membrane is calculated by the method of molecular dynamics (MD)., However, these studies did not refer the ion-dipole binding energy with phospholipid dipole heads. It would therefore be interesting to compare the results of the model [48] with the calculations performed by the MD. .

## V. Blocking and initiation of the action potential [50]

As shown in the preceding Sections, the ultrasound can lead to a redistribution of sodium channels. As a result, the regions in the axon membrane of high density and low density of the transmembrane sodium channels may be formed. Since the generation of an action potential in a neuron happens in the initial segment, non-myelinated part of the axon between the axon hillock and the first myelinated section, all the calculations below were done on the basis of the standard model of Hodgkin and Huxley for the squid axon [75]. This model was chosen because it is well studied. The experiments on self-excitation or blocking of the action potential propagation in the axon of the squid can be relatively easily carried out and the squid's axon is non-myelinated. That is, the numerical and experimental results will be qualitatively valid also for initial segments in neurons. Since the system of equations of Hodgkin and Huxley is well known, we do not present it here. Note that all the coefficients in the presented model calculations are taken for the temperature $T = 6.3^0 C$.

The redistribution of sodium channels was specified in calculations by the model function (Figure 12), so that the distribution of the sodium surface conductivity is $G_{Na} = G_{Na}^0 \cdot (1 + \xi_{Na})$. The length of the model axon was equal $l_a = 40$ cm.

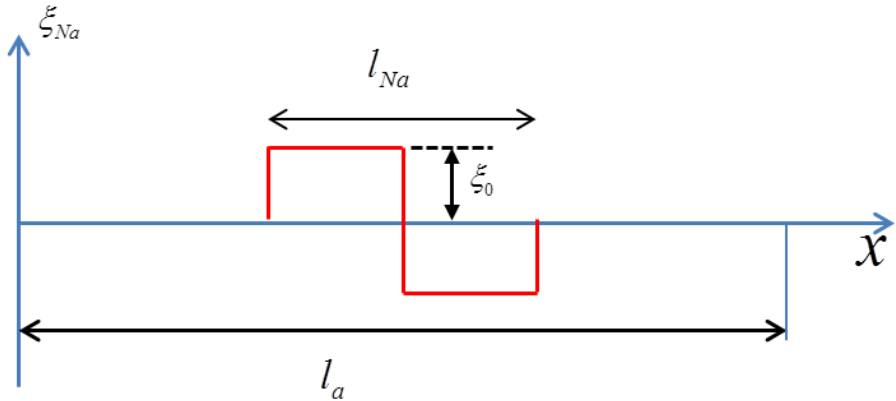

**Figure 12.** The model function $\xi_{Na}$ and the size of the redistribution channels, $l_{Na}$.

The excitation of the axon in the model was determined by a local increase in the potential in the point $x_0 = l_a / 2$ for the time interval $\tau_0 = 1$ms:

$$V_0(x_0) = V_R + \delta V \theta(t - \tau_0) = V_R + \alpha |V_R| \theta(t - \tau_0), \tag{40}$$

where $\theta(t - \tau_0) = \begin{cases} 1, & t \leq \tau_0 \\ 0, & t > \tau_0 \end{cases}$ is the step function.

Here parameter $\alpha$ characterizes the amplitude of perturbation of action potential. When $\alpha = 0$, the perturbing potential $\delta V = 0$ and the potential at the point $x_0$ is equal to the unperturbed potential of the membrane. The dependence parameter $\alpha$ on $\xi_0$ ($l_{Na} = 9$ cm, $x_0 = l_a / 2 = 20$ cm) is shown on Figure 13.

It is seen that when increasing the parameter $\xi_0$, the threshold value for the action potential excitation decreases, and the self-excitation takes place in the axon at $\xi_0 = 0.331$.

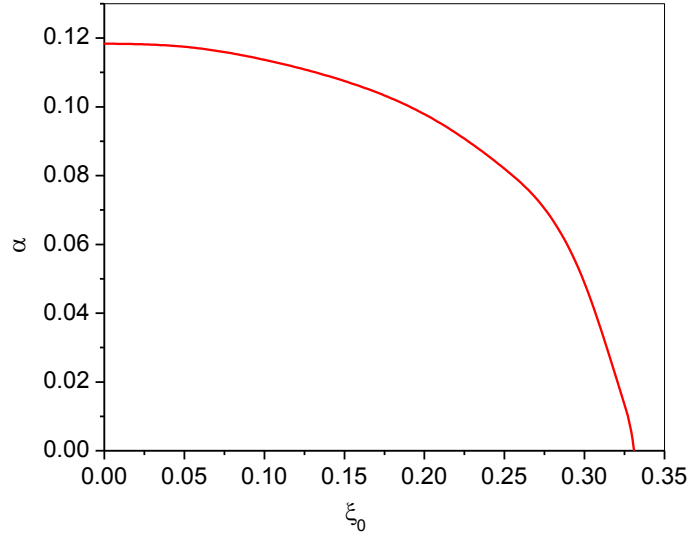

**Figure 13.** Dependence of the parameter $\alpha$ on $\xi_0$ at $l_{Na} = 9$ cm, $x_0 = 20$ cm.

By increasing the parameter $\xi_0$, the conditions of propagation to the right, i.e. in the region of low density of ion channels in the signal, deteriorates, and at $\xi_0 = 0.787$, the action potential does not extend to the right, e.g., in this direction, the blocking of the axon takes place. Figure 14 shows the dependence of the action potential on time at $\xi_0 = 0.81$. In this case, at $x = 20$cm the spontaneous excitation of the action potential propagating to the left occurs. In this example, the signal does not propagate to the right due to blockage of the axon.

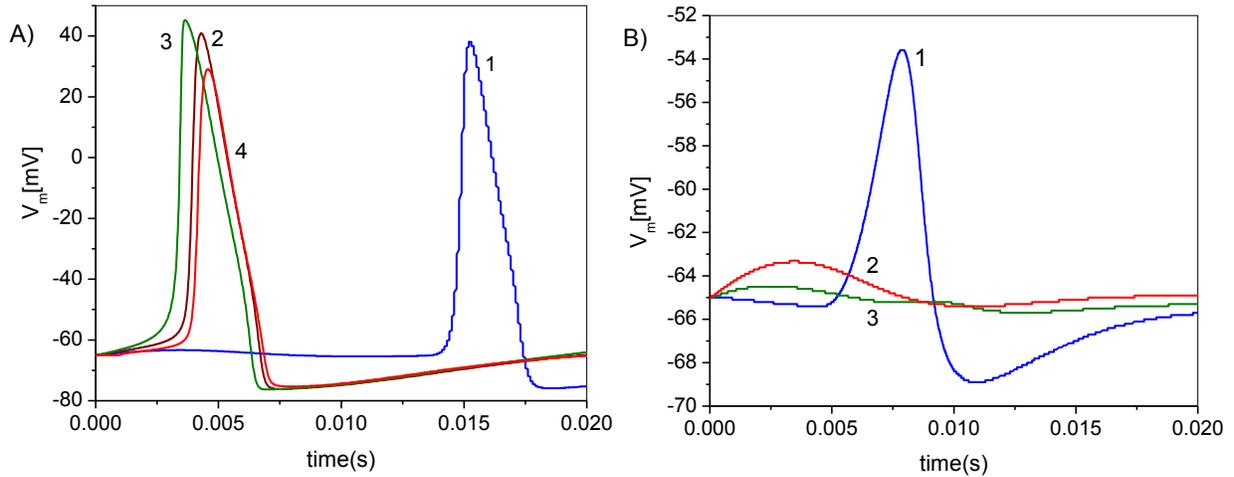

**Figure 14.** Time dependence of the action potential at various points in the axon. The excitation potential perturbation is $\delta V_m = 0$, $\xi_0 = 0.81$, $\alpha = 0$. A) Corresponds to points lying to the left of the center of the axon (where the action potential is excited), line 1 – corresponds to x=2cm, 2 – 15.5cm, 3 – 17.5 cm, 4 – 20 cm. B) Corresponds to points lying to the right of the center of the axon. 1- x=22.5cm, 2 –24.5cm, 3 – 30cm.

Figure 15 shows the dependence of the coefficient $\xi_0$, at which the action potential propagation is blocked, on the size of the region where the protein channels redistribution takes place, $l_{Na}$. These results correspond to the perturbation amplitude of the resting potential $\delta V_m = 0$ (parameter $\alpha = 0$).

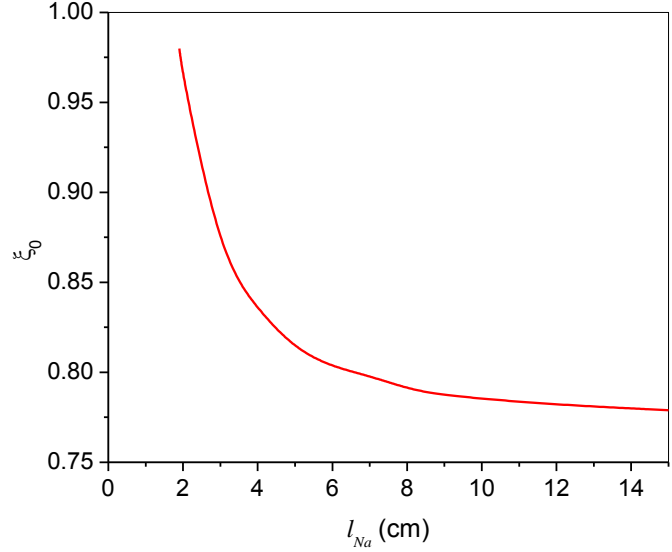

**Figure 15.** The dependence of the coefficient $\xi_0$ in the axon of squid on the size of the protein channels redistribution region $l_{Na}$ at $\alpha = 0$

In [50] it is shown that for ultrasonic acoustic waves, both standing and traveling, their intensities, at which effects of redistribution of ions channels can be observed in the membrane of the axon of squid, are very high and leads to the irreversible change in the membrane properties, related to the membrane heating and the cavitation development in the surrounding liquid, are occurring at much lower intensities of ultrasound. As regards the effects of the microwave exposure, then all estimates for squid axon are apparently greatly overestimated and were presented only to illustrate the effect, since the microwave radiation cannot be considered spatially uniform at the characteristic scales of squid axon, as it takes place for the case of AIS.

## VI. Discussion

Let us check that the linear elasticity theory is valid for the description of the forced vibrations of the membrane in a microwave field. Substituting (19) into (29), we obtain the expression for the amplitude of the displacement of the membrane for the first resonant harmonic:

$$A_\| = \frac{2}{\pi^2} \frac{\sigma_m}{\omega_r v h \rho_0} L_s^2 \sqrt{\frac{2 I_{MW}}{\varepsilon_0 \varepsilon_m^{1/2} c}} \tag{41}$$

$$A_\perp = -\frac{c_l k}{\omega_r} A_\| = -\frac{h}{L_s} A_\| \tag{42}$$

Since the longitudinal displacement of the membrane $A_\parallel$ should be much less than $L_s$, and the transversal displacement, $A_\perp$, is much less than $h$,

$$K_I = \frac{A_\parallel}{L_s} = \frac{A_\perp}{h} = \frac{2}{\pi^3} \frac{\sigma_m}{v c_s \rho_0 \sqrt{\varepsilon_0 \varepsilon_m^{1/2} c}} L_s \sqrt{2 I_{MW}} \approx 1.8 \cdot 10^{-3} L_s \sqrt{I_{MW}} \ll 1 \tag{43}$$

The constraint (43) imposes a condition on the microwave intensity. In the calculation of the constant in (43) we used the same parameters of the membrane, as in Section V. Following (39) we obtain:

$$\Delta t_s \approx 1.3 \cdot 10^{-5} \frac{1}{L_s^2 \cdot I_{MW}} \tag{44}$$

In the calculation of the constants in (43), (46) we used the same parameters of the membrane, as in Section V.

For a real initial segment length $L_s \approx 50\,\mu\text{m} \ll \lambda$ at the resonant microwave radiation with intensity $I_{MW} = 1\,\text{Watt/m}^2$, we obtain respectively $K_I \approx 9 \cdot 10^{-8}$ (i.e., elongation of the membrane is small enough to have been valid for the linear elasticity theory), and $\Delta t_s \approx 5.4 \cdot 10^3$ s. Corresponding energy density of induced resonant standing wave is quite small: $\bar{E} = \frac{1}{2} \rho \omega_r^2 A_\parallel^2 \approx 5.7 \cdot 10^3$ J/m$^3$.

As shown in this study, the redistribution of transmembrane channels, caused by standing or traveling ultrasonic waves in the membrane of the initial segment, leads initially to a decrease in the excitation threshold of nerve impulses, And then, with increasing amplitude of the channel density perturbations, to the possibility of spontaneous excitation of the action potential. However, the effect of purely mechanical ultrasonic vibrations is not as effective as the action of the forced electromechanical oscillation at the resonant frequencies, excited by the interaction of charged membranes with the microwave. Further increases in the amplitude of the transmembrane channel density perturbations can block the propagation of a pulse in areas with rarefied density of channels. This effect was likely observed in experiments [21], where the frog nerves were irradiated for 70 s with the microwave pulse of the frequency 60.125 GHz at a very weak intensity 70–770 nW/cm2. At the minimum microwave power, a spontaneous excitation of the action potential was observed. With increasing microwave power, the spontaneous excitation stopped and occurred again after switching off power. These results are fully qualitatively consistent with our model. In [21], a change in the shape of the action potential and shortening of the duration of the refractive period with increasing of the microwave power was also observed. All these experimental effects may be associated with the density variation of the transmembrane ion channels, induced by ultrasound [49]. It should be noted that the microwave frequency 60.125 GHz, used in [21], is likely to be close to the resonance frequencies of longitudinal vibrations of the axon membrane [49].

This raises the following question: how for effective interaction of charged particles attached to the membrane with the external field, the value of the field (the intensity of the microwave) may be very small. Regardless of the magnitude of the field, the energy and the momentum are transferred from the field to the charged particles. Thus, the random thermal motion is superimposed on a small ordered motion. This occurs, for example, in an electrolyte (plasma), when the drift velocity of the charged particles is a million times less than their average thermal velocities. The microwave field is interacting with the charges on the membrane and transmits the momentum to them, which, in turn, is transmitted to the membrane and causes forced oscillations. The same holds for a weak ultrasonic wave interacting with suspended particles (protein channels). As a result, an acoustic drift appears, superimposed on the average thermal motion, even at an ultrasound with a very low intensity. In this case, the potential of the acoustic force acting on the particles can be much smaller than the thermal energy, as it happens in the hydrodynamic or gas flows where the directional velocity is much smaller than the averaged thermal velocity, and the corresponding pressure drop is much lower than the unperturbed static pressure.

It should be noted that the resonant frequencies [32, 33], at which the displacement of the membrane is maximal, is proportional to the speed of sound in the membrane. In our estimations, we chose $c_s = 1500$ m/s, equal to the velocity of sound in the oil. It is difficult to know the exact value of the longitudinal speed of sound in the membrane at high frequencies, which could be much lower [76]. Since the resonance is very narrow [21], the experimentally measured frequencies, at which there is a blocking or self-excitation of the action potential, enable the determination of the velocity of sound in the membrane. It appears that the best experimental methods to observe the effect of transmembrane channel redistribution described in this paper and in [49], correspond to the measurements of the spatial distribution of sodium channels along the axon membrane, including electro-optical measurements (see [77, 78]), since they allow one to directly measure the density distribution of the transmembrane channels prior to the ultrasound (or microwave) pulse and immediately after. An indirect confirmation of transmembrane channel redistribution would be the observation of the action potential excitation and propagation (or blocking) dependence on the amplitude of the ultrasound. This can be done using conventional probe techniques [79], as well as nonintrusive optical methods, e.g., observing the dynamics of the second-harmonic generation of laser radiation scattered by the nerve fibers [80,81].

Let us formulate the primary experimental tasks that arise in the light of the proposed mechanism of interaction of low intensity microwave with membranes of nerve fibers.

1. **The redistribution of channels.** It is necessary to obtain data on the redistribution of transmembrane ion channels in the AIS or nemielirovannyh terminal fibers (Figure 6) when exposed to low intensity microwave.
2. **Measurement of the sound velocity in the membrane.** From the observation of the resonance frequencies of the microwave interaction with membrane, the speed of sound can determined, because of $\omega_r \propto c_l$.

## Conclusions

(1) The non-thermal microwave field can cause the ultrasonic mechanical vibrations in the nerve fibers.

(2) The presence of the resonances in the range of tens of GHz is in good agreement with known experimental data. Resonant frequencies are different for different membranes. This explains the contradictory experimental results obtained by different groups that investigated the influences of weak microwave radiation on cells. In our opinion, the success of the experiments4 is associated with the chosen frequency of microwave radiation 61.125 GHz, which is apparently close to one of the mechanical resonances of the nerve membrane. Furthermore, this frequency is close to the first resonance of forced longitudinal oscillations (74.6 GHz) obtained in the present study.

(3) The interaction with the ultrasound vibrations could change the density of transmembrane protein sodium channels. As a result, it could also change the resting potential in the nerve fibers and, therefore, it may induce or, conversely, suppress the excitation of the action potentials.

(4) The most effective component of the microwave electric field is the tangential component to the surface of the membrane, and the most effective region in the myelined axon is the initial segment, e.g., the section between the neuron and the first portion covered with a myelin sheath.

(5) Despite the encouraging results, our analysis is preliminary and requires further in-depth experimental and theoretical study. First of all, this analysis concerns with issues related to the density of the surface charges, the longitudinal and transverse sound velocities of in the membrane and the surrounding fluid, and viscous losses at GHz frequencies range, where the amplitude of ultrasonic vibrations is comparable to, or less than, the intermolecular distance in the liquid and membrane.

(6) We considered only the influence of ultrasonic standing waves excited by microwave radiation on the distribution of $Na^+$ transmembrane protein channels. However, similar effects should be expected for other kinds of membrane proteins (such as transmembrane ion channels, as well as the peripheral surface proteins) affected by lateral diffusion. This, in principle, could significantly increase the effects caused in the nerve fibers by the weak microwave radiation.